# Surface Structures of Epitaxial B20 FeGe($\bar{1}\bar{1}\bar{1}$) Thin Films via Scanning Tunneling Microscopy

J. P. Corbett,[1] T. Zhu,[1] A. S. Ahmed,[1] S. J. Tjung,[1] J. J. Repicky,[1] T. Takeuchi,[2] R. K. Kawakami,[1] and J. A. Gupta[1,a)]

[1] *Department of Physics, the Ohio State University, Columbus, OH 43210, USA*

[2] *Integrated Graduate School of Medicine, Engineering, and Agricultural Science, University of Yamanashi, Kofu, 400-*

We grew 20-100 nm thick films of B20 FeGe by molecular beam epitaxy and investigated the surface structures via scanning tunneling microscopy. We observed the atomic resolution of each of the four possible chemical layers in FeGe($\bar{1}\bar{1}\bar{1}$). An average hexagonal surface unit cell is observed via scanning tunneling microscopy, low energy electron diffraction, and reflection high energy electron diffraction resulting in a size of ~6.84 Å in agreement with the bulk expectation. Furthermore, the atomic resolution and registry across triple-layer step edges definitively determine the grain orientation as (111) or ($\bar{1}\bar{1}\bar{1}$). Further verification of the grain orientation is made by Ar+ sputtering FeGe($\bar{1}\bar{1}\bar{1}$) surface allowing direct imaging of the subsurface layer.

## I. INTRODUCTION

FeGe can crystalize into three polymorphs, a hexagonal phase, a monoclinic phase, and the cubic B20 phase[1] of interest here. The B20 family of materials (e.g., FeGe, MnSi, PdGa, AlPd, MnGe) have the interesting property of breaking inversion symmetry along the {111} planes which can lead to magnetic skyrmions[2,3], topologically protected spin textures[4]. Skyrmion sizes have been reported down to nanometer scales[5–7], which is attractive for high density storage[4,5,8]. Though B20 materials host skyrmions in the bulk, there is interest in adding interfacial tuning to further decrease the size and increase critical temperature ($T_C$) above room temperature[9]. The sensitivity of these effects to interface quality and termination motivates the study with atomic resolution imaging techniques including scanning tunneling microscopy (STM) and transmission electron microscopy (TEM), as well as spin sensitive extensions (i.e., spin-polarized STM and Lorentz TEM).

Previous STM studies on B20 MnSi have revealed the characteristic layered structure, and the coexistence of multiple surface terminations corresponding to sparse[10] and dense[11,12] atomic layers in the B20 structures. These terminations were evident as terraces with distinct levels of corrugation in atomic resolution images, separated by varying fractions of the characteristic B20 quadruple layer (QL) structure[11,10,12]. Density functional theory (DFT) of MnSi and AlPd predicted both spare and dense layers as the lowest energy chemical terminations for both orientations of the (111) or ($\bar{1}\bar{1}\bar{1}$)[10,13]. Comparison of the experimental data with DFT calculations helped unravel the complex surface structure of MnSi including adatom features[10].

Here, we report the first STM study of B20-phase FeGe. Epitaxial thin films are grown by molecular beam epitaxy (MBE). We observe four distinct levels of corrugation in atomic resolution images of the FeGe($\bar{1}\bar{1}\bar{1}$), which we denote as very high corrugation (vHC), very low corrugation (vLC), high corrugation (HC), and low corrugation (LC). Thorough analysis of the surface structure registry across terrace steps and comparing tunneling spectroscopy with recent theory calculations[14], we assign the four corrugations vHC, HC, LC, and vLC to the Fe-dense (Fe-d), Fe-sparse (Fe-S), Ge-d (Ge-d), and Ge-sparse (Ge-s) surfaces, respectively. These represent the four possible chemical terminations within the quadruple layer stacking sequence in FeGe. By resolving the stacking sequence across fractional QL steps and in partially exposed surfaces via Ar+ sputtering, we are able to distinguish between (111) or ($\bar{1}\bar{1}\bar{1}$) orientations, which is difficult by conventional techniques such as x-ray diffraction since the interatomic spacing of both orientations are equal. This capability is important for SP-STM imaging of skyrmions, as the orientation of the grain determines the chirality of the skyrmion. Additionally, knowledge of the chirality and terminating layer is of importance for the interfacial tuning of B20 superlattices.

## II. EXPERIMENTAL

Epitaxial growth of FeGe thin films is achieved using a custom-built MBE system with a base pressure of $\sim 1.0 \times 10^{-10}$ Torr. The growth is monitored in real-time with a 10 KeV reflection high energy electron diffraction (RHEED) system from STAIB Instruments, while the fluxes are calibrated using a quartz crystal thickness monitor. The Si(111) substrates are prepared by first precleaning the substrate ultrasonically with acetone, then isopropanol for 5 minutes each. The Si(111) substrate is subsequently dipped in a buffered HF solution for 2 minutes to chemically remove the oxide layer and H terminate the Si(111) surface. The substrates are immediately transferred into the MBE chamber via a load lock to prevent re-oxidation. The Si(111) substrates are then annealed at 800 °C for 20 minutes until RHEED shows a streaky $7 \times 7$ reconstruction pattern. The substrate is then cooled to the growth temperature of 300 °C and Fe and Ge are co-deposited using thermal effusion cells with fluxes matched 1:1 in order to achieve a stoichiometric film. Films were grown with 20-100 nm thickness; all data reported here were on the 20 nm films, but were consistent across samples. Further details and additional characterizations of the growth are reported elsewhere[9].

Shortly after growth, the FeGe samples are transferred *in-situ* into a Ferrovac UHV vacuum suitcase with a base pressure of $\sim 1.0 \times 10^{-9}$ Torr and transferred to another lab housing a Createc LT-STM system with its own UHV load lock and analysis chambers. LEED patterns were collected at normal incidence with a 140 eV beam energy to confirm crystallinity of the films. FeGe samples were directly transferred into the cold STM without any additional surface preparation to minimize contamination and preserve the as-grown film quality. STM measurements were performed at 5K using cut PtIr tips. Tunneling spectroscopy was performed by adding a 20mV modulation voltage to the DC bias. Image processing and analysis is preformed using Gwyddion and WxSM for RHEED, LEED and STM data.[15,16]

## III. RESULTS AND DISCUSSION

From a growth perspective, breaking the inversion symmetry of the {111} planes leads to two possible stacking directions, along [111] or [$\bar{1}\bar{1}\bar{1}$]. The QL structure in FeGe($\bar{1}\bar{1}\bar{1}$) is shown in FIG. 1 where a chemical sequence of Fe-S, Ge-S, Fe-d, and Ge-d layers repeats three times (3 QL) before returning to the same relative atomic positions, thus resulting in a 12 layer unit cell. Sparse layers correspond to monomer Fe, Ge in a hexagonal lattice, while dense layers correspond to trimers of Fe, Ge arranged in the same lattice. In FeGe(111) the stacking also follows a repeating 3QL pattern, but with a different chemical sequence (Fe-s, Ge-d, Fe-d, and Ge-s). We present below a detailed analysis which helps distinguish between these two grain orientations in our STM images.

### *Morphology and Surface Unit Cell Determination*

**Error! Reference source not found.** Shows the LEED and RHEED images of the as-grown FeGe surface. The 'streaky' RHEED pattern indicates high quality, uniform film growth. The LEED pattern was taken after the UHV vacuum suitcase transfer from the MBE system to the STM system, which demonstrates that the suitcase transfer does not add significant surface contaminants. LEED shows a hexagonal pattern, which is in agreement with the expected symmetry of FeGe{111} (c.f., top views in FIG. 1). We measure surface lattice constants of 6.6 $\pm$ 0.50 Å (LEED) and 6.74 $\pm$ 0.50 Å, (RHEED) which are in good agreement with bulk FeGe crystals (6.64 Å)[9].

From the RHEED and LEED measurements we expect to see large flat terraces with a hexagonal atomic structure terminating the surface. **Error! Reference source not found.**(a) shows an STM image of the surface, indeed showing several large, atomically flat terraces. The shape and edge angles of the terraces match the symmetry of a (111) film orientation with edges meeting at 60° and 120° angles. As expected from the LEED pattern, atomic resolution images of these terraces (FIG. 3a, inset) reveals a hexagonal lattice with a surface unit cell length of 6.91 $\pm$ 0.14 Å, within uncertainty of the RHEED and LEED surface unit cell measurements. The apparent step height between each terrace in FIG. 0 measures 2.77 ± 0.23 Å (FIG. 3b), in good agreement with x-ray diffraction measurements of 2.703 Å for the FeGe(111) QL.[9]

## *Determination of surface terminations*

Figures 4(a,e) show atomic resolution STM images of the FeGe surface which reveal regions with four distinct atomic corrugations. Bright features in all regions define a hexagonal lattice with a lattice constant of 6.9 Å. Figures 4(a-b) shows adjacent terraces of 'high corrugation' (HC) and 'low corrugation' (LC) surfaces, exhibiting ~ 20pm and ~10pm corrugation respectively in the atomic resolution images. Figure 4(e-f) shows an STM image of adjacent terraces of two additional distinct regions, we denote by 'very high' and 'very low' corrugations of ~ 40pm (vHC) and ~ 6pm (vLC) respectively. From our survey of the FeGe(111) surface the LC and vLC corrugations are predominately observed, with small regions of neighboring HC and vHC respectively.

Because these regions are imaged simultaneously, these differences cannot be attributed to the tip sharpness; we instead suggest that they represent distinct chemical terminations. Analysis of the step heights between these distinct regions is an initial step toward the chemical assignments. We note that in general there are topographic and electronic contributions to the apparent height in STM measurements. A true topographic height difference is measured between two terraces if they are chemically identical. For example, the apparent height difference between two adjacent HC regions is 2.7 ± 0.2 Å, consistent with the FeGe QL as shown in FIG. 3. Fractions of a QL however, are observed between HC and LC regions, and are indicated in FIG. 4a. For example, starting from the LC region in the middle of FIG. 4a, there is a 0.81 ± 0.15 Å step down to the HC region in the top center, and a 1.92 ± 0.10 Å angstrom step up to the HC region on the left. These are within uncertainty to the expected height differences for mono (ML) and triple layer (TL) steps in the FeGe B20 structure respectively (c.f., FIG. 1). Since a partial QL step moves between different terminations, the apparent step height is expected to be a mixture of electronic and topographic contributions. For the +1 V scanning bias used during imaging, the electronic contribution to the topographical signal appears to be small given we observe the expected ML and TL steps. Similarly, we observe atomic QL steps between two adjacent vLC terraces, while fractional QL steps are observed between adjacent vLC and vHC terraces. In FIG. 4(e) for example, we observe a ML step down from the vLC region to a vHC region, followed by a TL step down to the next vLC region. Since the FeGe(111) structure alternates in chemical termination every odd # of layers, these step heights and distinct contrast suggest that these regions have distinct chemical terminations, with an atomic structure and local density of states that translates to distinct apparent corrugations in STM images.

To determine which chemical species corresponds to these regions, we performed tunneling spectroscopy shown in FIG. 4(c), with the tip positioned over bright spots in vHC and vLC regions as indicated in FIG. 4(e). A much larger dI/dV signal is observed on vHC regions, with two prominent broad peaks at ~-0.3 V and +0.2V visible as shoulders above the background, while spectroscopy on the LC regions reveals a smaller conductivity and is relatively flat and featureless.

This comparison is consistent with spatial mapping of the *dI/dV* signal at +1.0 V in FIG. 4(d). HC regions show a bright contrast (yellow), compared to the LC, which shows a dark contrast (purple). These data indicate that HC regions have a higher local density of states than LC regions. For comparison, Yamada et al. performed self-consistent linear-muffin tin orbital calculations on B20 FeGe showing peaks near the Fermi level predominantly derived from Fe states and in agreement with our observed peaks, while the DOS of Ge is low and flat[14].

We propose a structural model assigning the vHC (LC) corrugations to Fe-d (Ge-d) terminations respectively, and the HC (vLC) corrugations to Fe-s (Ge-s) terminations. Sparse layers correspond to monomer Fe,Ge in a hexagonal lattice, while dense layers correspond to trimers of Fe,Ge arranged in the same lattice. To confirm this picture, we show high resolution filled state images of the four surface terminations and corresponding atomic models in FIG. 5. Our results are in qualitative agreement with Tersoff-Hamann simulations of AlPd{111}, which showed similar contrast for Al-sparse and Pd-dense surface terminations as we discuss below[13].

For the Fe-s surface termination FIG. 5(a-b), Fe atoms (yellow in FIG. 5) are located in threefold hollow sites of an underlying Ge-s (blue in FIG. 5b) lattice, and as observed in the STM image in FIG. 5a) there is a triangular, medium contrast region at the geometric center of spherically symmetric protrusions, hinting at the underlying Ge-s lattice. The G-s surface termination appears as spherical protrusions sitting atop of the Fe-d sublattice FIG. 5(c-d). The Ge atoms in the Ge-s lattice sit atop of the medium-sized isosceles triangle of Fe atoms in the dense lattice. Similar hints of the sublattice are observed for the Ge-s termination, in this case arising from trimers in the underlying Fe-d layer (FIG. 5c-d). Discussed below is an STM image of a sputtered Fe-s surface, where this underlying lattice is resolved even more clearly.

A high-resolution STM image of the Fe-d termination and accompanying atomic model are shown in FIG. 5(e-f). In this surface, Fe trimers are arranged in a triangular lattice over a Ge-d sublattice. The triangular shape of the protrusions in the STM image is clearly distinct from the more spherical shape of the atoms in the Fe-s termination. The lateral size and orientation of the triangular protrusions agree well with the model's trimer size and orientation. Lastly, FIG. 5(g-h) shows the Ge-d termination. The protrusions do not exhibit triangular shape as in the Fe-d termination, which may be due to the smaller size of the Ge trimers and/or the lower DOS of Ge in the FeGE surface. Nevertheless, these features do exhibit more of an oblong shape compared to the more spherically symmetric Ge-s surface, which we attribute to the Ge-trimers in the dense termination.

As further evidence in our assignment of the FeGe surface terminations, FIG. 6 (a) shows an STM image of the $Ar^+$ sputtered FeGe($\bar{1}\bar{1}\bar{1}$) surface, where the surface termination layer was partly removed. The topmost atoms (orange) retain their hexagonal symmetry and even form complete patches of the surface unit cell, with a corrugation that matches the HC structure (FIG. 6(b)), indicating that this is the Fe-s termination. Where topmost Fe atoms have been removed, small medium-contrast (purple) protrusions can be seen which are arranged in triangular manner around the Fe-s atoms. FIG. 6(c) is a model of the Fe-s layer atop of a Ge-s layer, with selected Fe atoms removed from the surface structure to match the STM image. We find excellent registry of both the topmost Fe atoms, as well as the Ge-s subsurface layer.

We find the Fe-s and Ge-s terminations to dominate the surface, suggesting that these have the lowest surface energies under our growth conditions. The lowest surface energy depends on the growth composition (i.e. the flux ratio), where growth under Fe-deficit or Fe-rich conditions can impact the local surface structure. For MnSi the lowest energy structure of the (111) direction depends on the growth stoichiometry where the surface structure transitions from the Si-sparse termination to the Mn-dense termination with decreasing Si-richness[13]. We find Ge-d and Fe-d regions on the edges of incomplete QL terraces, indicating a local deficiency of Fe or Ge respectively in the terrace formation.

### *Chirality determination via atomic registry across QL and TL steps*

An interesting, and important point is that FeGe(111) is distinguishable from the FeGe($\bar{1}\bar{1}\bar{1}$) as these orientations have different stacking chirality. For ($\bar{1}\bar{1}\bar{1}$), we expect a stacking order of Fe-s→Ge-s→Fe-d→Ge-d, while for (111), we expect a different order of Fe-s→Ge-d→Fe-d→Ge-s. This sequence can be revealed in STM images of topographic areas showing terraces separated by incomplete QL steps (i.e., ML or TL), and is confirmed by looking at the registry of the lattices across these steps. From FIG. 1, we expect a shift in the atomic lattices between terraces separated by a QL. For ML steps, we expect a change in atomic contrast as well as a shift in lattice, while for TL steps, we expect a change in atomic contrast, but no shift in lattice. FIG. 7 shows Laplacian filtered STM images across QL and TL steps with accompanying atomic models to confirm this registry. FIG. 7(a) shows a QL step (indicated by the blue dashed line) between two Fe-s terraces. Three white lines are overlaid in the STM image and aligned to the hexagonal structure on the bottom terrace. Extending these lines to the upper terrace, we see that the two structures align along the [11$\bar{2}$] direction (with an apparent shift of $\frac{\sqrt{3}}{6}a$) , but do not align along the [10$\bar{1}$] or [01$\bar{1}$] directions. This behavior is reproduced in the corresponding atomic model of the Fe($\bar{1}\bar{1}\bar{1}$) surface, showing the Fe-s termination atop a Ge-s layer (FIG. 7FIG. 5b). The model has excellent agreement with the experimental image, as the two lattices align along the [11$\bar{2}$] direction with the apparent shift of $\frac{\sqrt{3}}{6}a$, but do not align along the $[10\bar{1}]$ or $[01\bar{1}]$ directions. Likewise, in FIG. 7(c) a QL step between two Ge-s terraces shows the same registry agreement and apparent shift as anticipated from the crystal structure of FeGe, and is in agreement with the structural model.

FIG. 7(e-f) shows a Laplacian filtered STM image with a TL step up in topography from a Ge-s region to a Fe-d region and the accompanying atomic model. The registry of the Ge-s to Fe-d layers as well as the lateral size and orientation of the Fe trimers are in excellent agreement between the experimental image and the structural model. The three white lines overlaid in FIG. 7(e) are aligned to the hexagonal structure of the Ge-s lattice and extended onto the Fe-d lattice. Both lattices align along all three directions and show no apparent shift which is expected from the alignment of the Ge-s to Fe-d layer across a TL step. FIG. 7(g-h) shows an STM image with a TL step down from a Fe-s surface to a Ge-d surface and the accompanying structural model. The registry and lack of apparent shift in the two lattices is in good agreement.

From our observation of partial QL steps and the sputtered surfaces we find we only imagined FeGe($\bar{1}\bar{1}\bar{1}$) grains, but our analysis enabled us to definitively determine the chirality which is a key ingredient in the interpretation of spin-polarized STM imaging of skyrmions.

### III. CONCLUSIONS

We observe atomic resolution of each of the four possible chemical layers in our FeGe($\bar{1}\bar{1}\bar{1}$) MBE prepared thin films. The four surface structures all have the same surface unit cell length of in agreement with complimentary LEED and RHEED measurements given an average unit cell length of ~6.84 Å. Moreover, the atomic resolution and registry across a TL step edges enables definitive determination of the grain orientation as (111) or ($\bar{1}\bar{1}\bar{1}$) planes. As an additional verification of the grain orientation, Ar+ sputtering the surface enables direct imaging of the subsurface layer to confirm grain orientation.

**ACKNOWLEDGEMENTS**

We acknowledge the support of from the Defense Advanced Research Projects Agency (Grant No. D18AP00008) The authors thank R. K. Smith for useful discussions.

**REFERENCES**

1. Zeng, C. *et al.* Epitaxial stabilization of ferromagnetism in the nanophase of FeGe. *Phys. Rev. Lett.* **96,** 2–5 (2006).

2. Zhao, X. *et al.* Direct imaging of magnetic field-driven transitions of skyrmion cluster states in FeGe nanodisks. *Proc. Natl. Acad. Sci.* **113,** 4918–4923 (2016).

3. Turgut, E. *et al.* Engineering Dzyaloshinskii-Moriya interaction in B20 thin film chiral magnets. *arXiv* 3–5 (2018).

4. Wiesendanger, R. Nanoscale magnetic skyrmions in metallic films and multilayers: A new twist for spintronics. *Nat. Rev. Mater.* **1,** (2016).

5. Jiang, W. *et al.* Blowing magnetic skyrmion bubbles. *Science (80-. ).* **349,** 283–286 (2015).

6. Tanigaki, T. *et al.* Real-Space Observation of Short-Period Cubic Lattice of Skyrmions in MnGe. *Nano Lett.* **15,** 5438–5442 (2015).

7. Grenz, J., Köhler, A., Schwarz, A. & Wiesendanger, R. Probing the Nano-Skyrmion Lattice on Fe/Ir(111) with Magnetic Exchange Force Microscopy. *Phys. Rev. Lett.* **119,** 1–4 (2017).

8. Fert, A., Cros, V. & Sampaio, J. Skyrmions on the track. *Nat. Nanotechnol.* **8,** 152–156 (2013).

9. Ahmed, A. S., Esser, B. D., Rowland, J., McComb, D. W. & Kawakami, R. K. Molecular beam epitaxy growth of [CrGe/MnGe/FeGe] superlattices: Toward artificial B20 skyrmion materials with tunable interactions. *J. Cryst. Growth* **467,** 38–46 (2017).

10. Geisler, B. *et al.* Growth mode and atomic structure of MnSi thin films on Si(111). *Phys. Rev. B - Condens. Matter Mater. Phys.* **86,** 1–7 (2012).

11. Azatyan, S. G., Utas, O. A., Denisov, N. V., Zotov, A. V. & Saranin, A. A. Variable termination of MnSi/Si(111)$\sqrt{3}\times\sqrt{3}$ films and its effect on surface properties. *Surf. Sci.* **605,** 289–295 (2011).

12. Suzuki, T. *et al.* Surface morphology of MnSi thin films grown on Si(111). *Surf. Sci.* **617,** 106–112 (2013).

13. Krajčí, M. & Hafner, J. Surfaces of intermetallic compounds: An ab initio DFT study for B20-type AlPd. *Phys. Rev. B - Condens. Matter Mater. Phys.* **87,** 1–15 (2013).

14. Yamada, H., Terao, K., Ohta, H. & Kulatov, E. Electronic structure and magnetism of FeGe with B20-type structure. *Phys. B Condens. Matter* **329–333,** 1131–1133 (2003).

15. Nečas, D. & Klapetek, P. Gwyddion: An open-source software for SPM data analysis. *Cent. Eur. J. Phys.* **10,** 181–188 (2012).

16. Horcas, I. *et al.* WSXM: A software for scanning probe microscopy and a tool for nanotechnology. *Rev. Sci. Instrum.* **78,** (2007).

17. Smith, A. R. Atomic-Scale Spin-Polarized Scanning Tunneling Microscopy and Atomic Force Microscopy: A Review. *J. Scanning Probe Microsc.* **1,** 3–20 (2006).

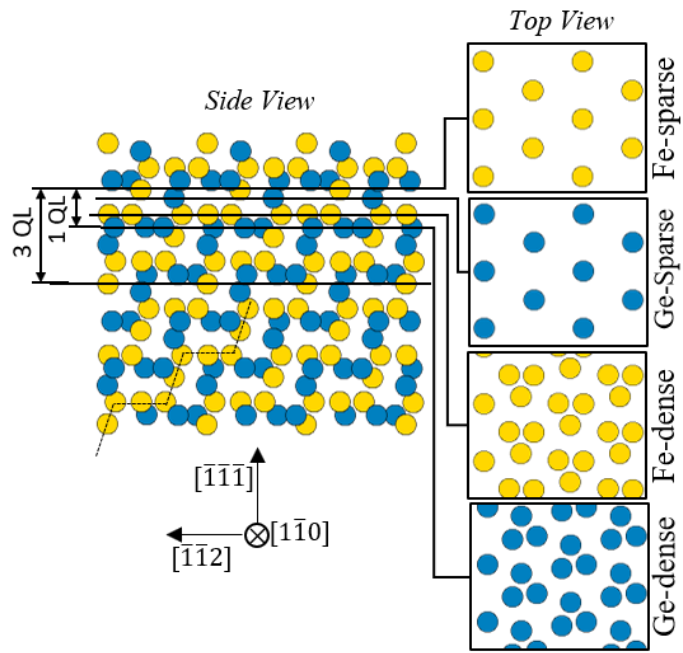

FIG. 1 Side view (left) of the model crystal structure of the FeGe($\overline{1}\overline{1}\overline{1}$). There are four distinct chemical layers in 1 QL of FeGe($\overline{1}\overline{1}\overline{1}$), a Fe-s, Ge-d, Fe-d, and Ge-s layer which are shown to the right in a top view. The dotted zig-zag line indicates the staircase structure of the FeGe{111} layers.

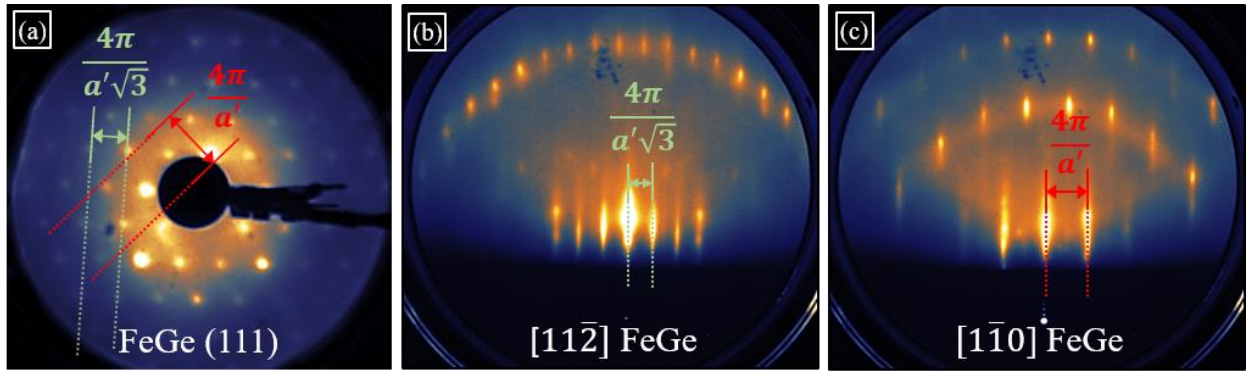

FIG. 2 LEED and RHEED patterns of the FeGe(111) surface. (a) LEED pattern taken at normal incidence with a 140 eV energy beam. The reciprocal lattice distances are indicated with green and red markers. (b-c) The RHEED patterns taken along the $[11\bar{2}]$ and $[1\bar{1}0]$ direction respectively with a 10 keV energy. The 0$^{th}$ order reciprocal spacings are labeled for each direction, with a red and green color correspondence to the equivalent LEED spot spacing.

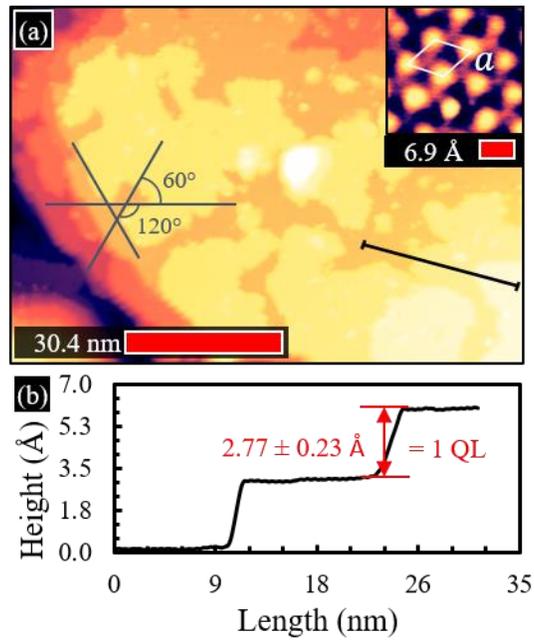

FIG. 3 STM image and line profile of the FeGe($\bar{1}\bar{1}\bar{1}$) surface (a) Large area STM image of an atomically smooth surface. The inset in the upper right corner is a high-resolution image of the hexagonal surface structure found on the terrace. The black line indicates a line profile across two step edges, which is shown in (b). (b) Line profile from (a) showing the step height to be a QL.  scanning conditions: $V = +1$ V, $I = 30$ pA.

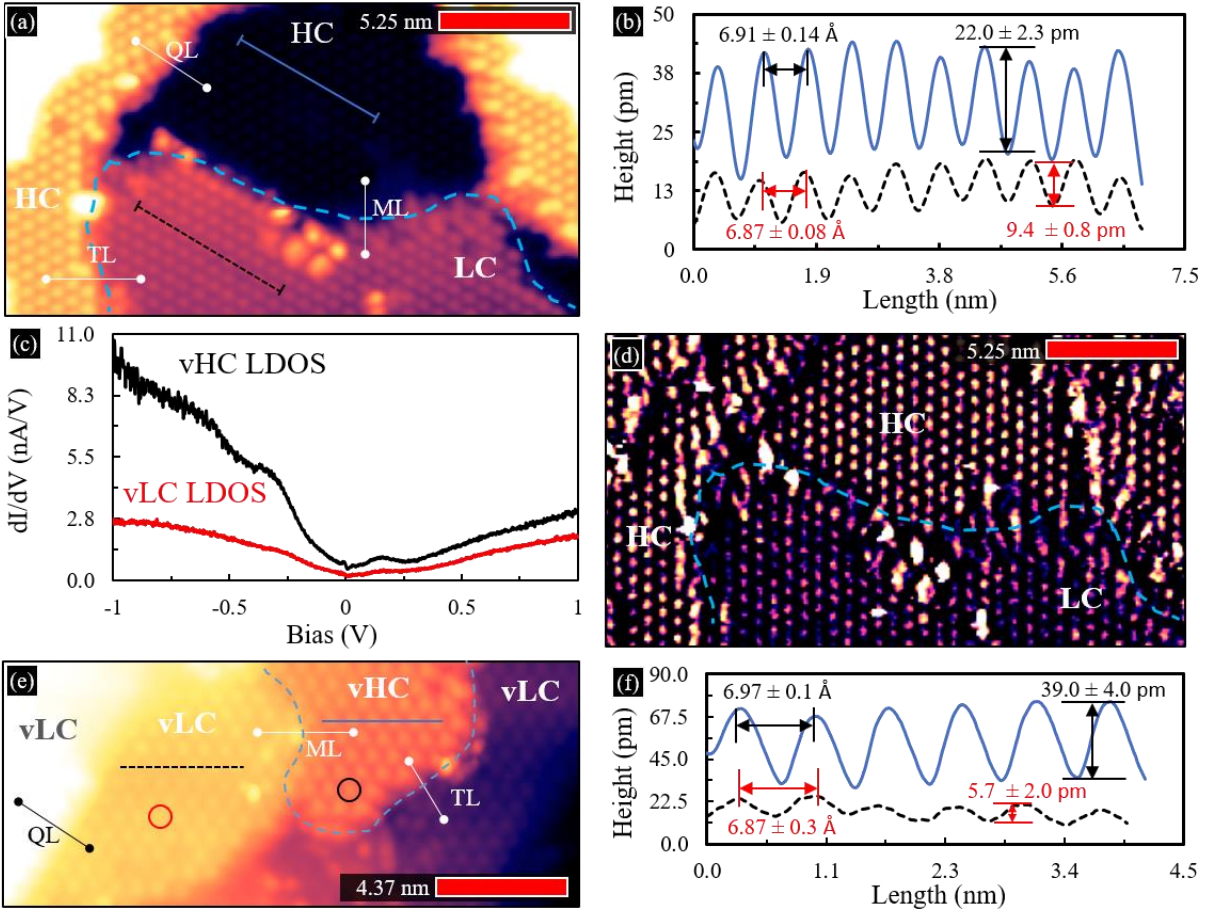

FIG. 4 STM images of the vHC, HC, LC, and vLC surface structures with corresponding line profiles and STS. (a) STM image of across a valley region with resolved atomic structures. The dashed blue line indicates the boundary between the HC and LC structures. We observe three types of apparent step heights a ML, TL, and QL. Line profiles of the LC (black dashed) and HC (blue solid) are shown in (b). (b) Line profiles of the HC and LC structures taken from (a).The apparent corrugation of the HC surface structure is ~2 as deep as the LC. The lattice size of the HC and LC are in agreement. (c) Point spectroscopy atop of the vHC atoms and vLC atoms from the image in part (e). The positions of the spectroscopy are mark with a red and black circles where the red cicle is for vLC and black circle for vHC. A significant increase in conductance on the vHC, and relatively flat and low conductance of the vLC tells us the vHC is formed of Fe atoms while the vLC is Ge. (d) dI/dV map of the HC and LC with the same boundary marked in (a). A contrast between HC (bright yellow) and LC (dark purple) is observed indicating a chemical difference between the two regions. Similar to part (c) a relatively higher conductance is measured for the HC versus the LC region suggesting HC is comprised of Fe and LC of Ge (e) A stepped region showing a ML, TL, and QL step with the vLC and vHC being observed. The blue dashed line indicates the boundary between the vLC and vHC stuctures. The line profiles of the vHC and the vLC structures are ploted in (f) where the solid blue line is across the vHC structure and the black dash is across the vLC.structure. (f) the line profiles of the vLC and vHC structures taken from (e). The vHC is ~10 larger than the vLC and ~2 the HC.  Scanning conditions: V = +1 V, I = 30- 50 pA.

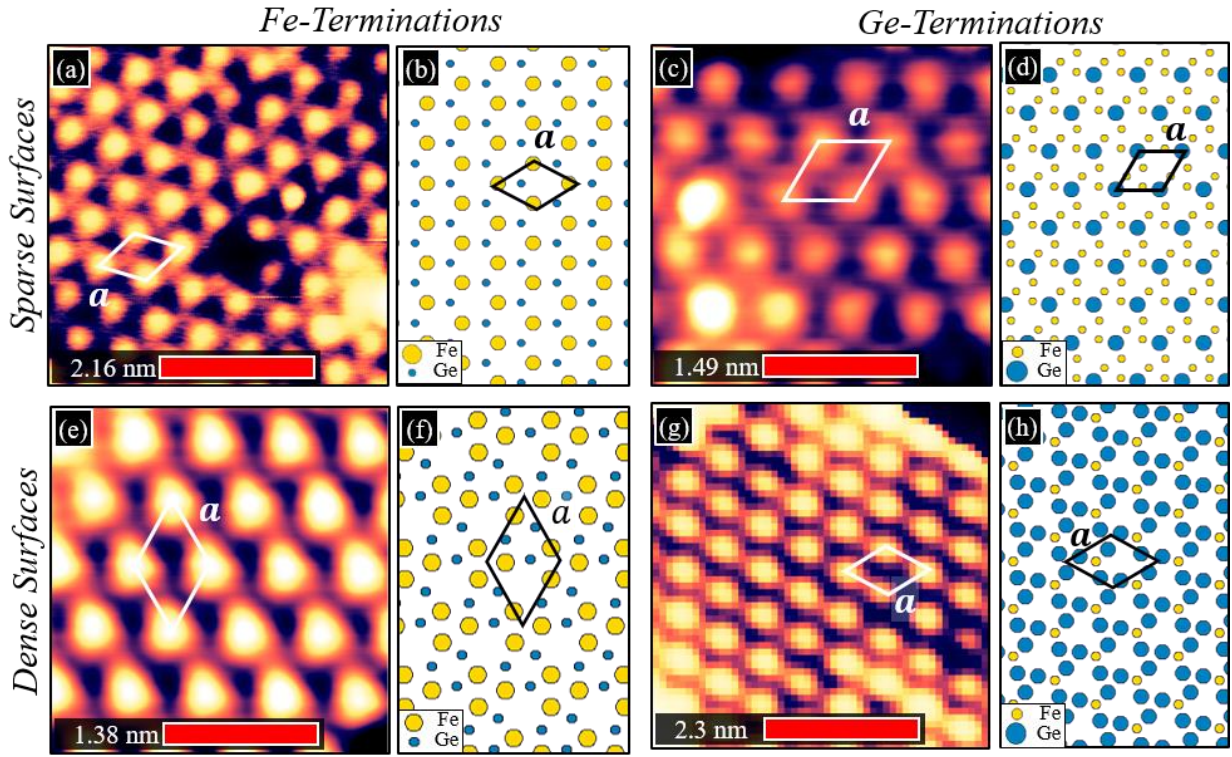

FIG. 5 STM images and to scale models of the four possible unreconstructed layersof the FeGe($\overline{1}\overline{1}\overline{1}$) stacking. (a) Filled state image of Fe-s termination, the bright (yellow) protrusions are individual Fe atoms arranged in a hexagonal unit cell. The inset in (a) is the empty state image of the upper right corner of the filled state STM image. The size scales are of (a) and the inset are identical. (b) Model of the Fe-s termination on a Ge-s subsurface. (c) Filled state imaging of the Fe-d termination. The bright triangular protrusions are Fe trimers arranged in a hexagonal unit cell. (d) Model of the Fe-d layer atop a Ge-d subsurface. (e) Filled state imaging of the Ge-s layer. The bright protrusions are individual Ge atoms in a hexagonal unit cell. (f) The model of the Ge-s layer atop of a Fe-d sublattice. (g) The filled state imaging of the Ge-d termination. The bright oblong protrusions are Ge-trimers. (h) The model of the Ge-d layer atop of an Fe-s subsurface layer. Filled sate scanning conditions: V = +1 V, I = 30- 50 pA. Empty state scanning condition. V = - 200 mV, I = 30 pA

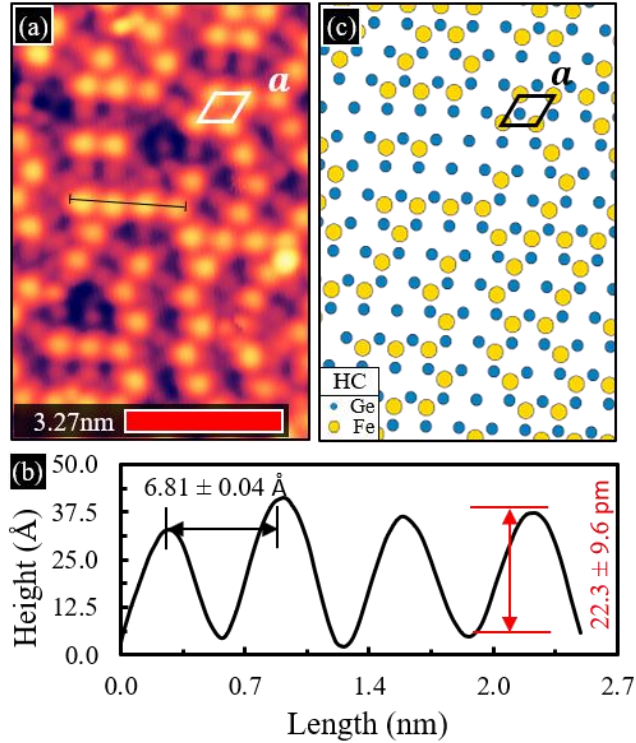

FIG. 6 Fill state STM image and surface model of the Ar+ sputtered Fe-s surface. (a) STM image of an Fe-s layer with significant amount of Fe atoms (orange protrusions) removed from the Ge-s subsurface layer. The Ge atoms can be seen as purple protusions between the orange Fe atoms. The white parrallegram indicates the unit mesh, which is still preserved despite the sputtering. The black solid line is the position of a line profile shown in (b). (b) Line profile from (a) showing the corrugation and atomic spacing to be that of the Fe-s layer. (c) Model of the Fe-s layer with selected Fe atoms (yellow circles) removed to match the surface. Scanning conditions: $V = 1$ V, $I = 30$ pA.

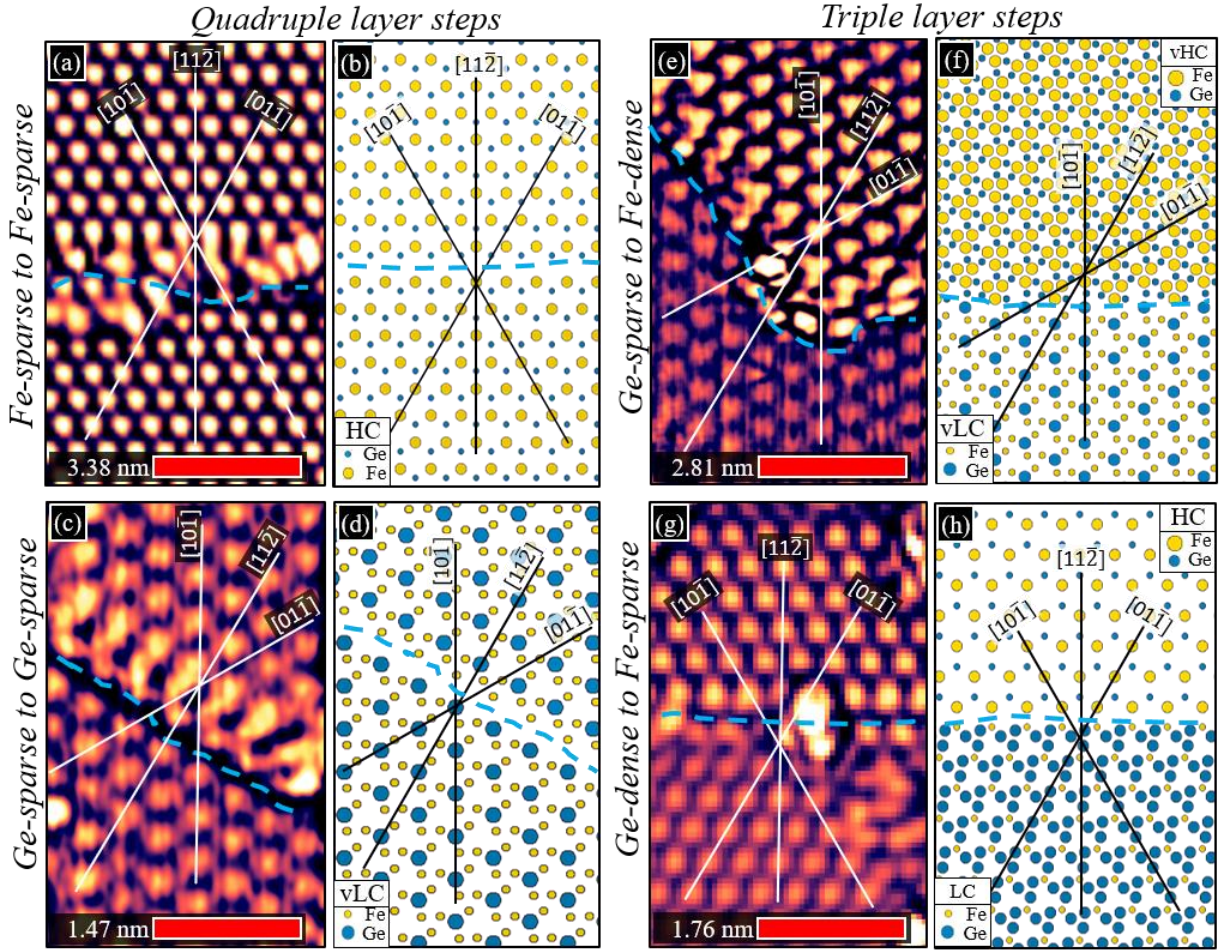

FIG. 7 STM images and models across step edges. The dashed blue line indicates a step edge, while the white and black solid lines are drawn to show relative atomic positions across a step edge. (a) STM image of the HC structure with a QL step edge running through the center. The white lines are aligned to the bottom region and show relative shift of the HC lattice after traversing a QL. (b) Model of the Fe-s to Fe-s termination separated by a QL drawn to the scale of (a).The solid black lines show the relative shift of the HC after stepping by a QL. (c) STM image of the vLC structure with a QL step edge diagonal through the center. The white lines are aligned to the bottom region and show relative shift of the vLC lattice after traversing a QL. (d) Model of the Ge-s to Ge-s termination separated by a QL drawn to the scale of (a).The solid black lines show the relative shift of the vLC after stepping by a QL. (e) STM image of the vLC structure separated by TL down to a vHC structure. The white line is aligned to the vHC and show no relative shift compared to the vLC. (h) Model of the Ge-s to Fe-d termination separated by a TL drawn to the scale of (e). (g) STM image of the HC structure separated by TL down to a LC structure. The white line is aligned to the HC and show no relative shift compared to the LC. (h) Model of the Fe-s to Ge-d termination separated by a TL drawn to the scale of (g). Scanning conditions: V = +1 V, I = 30- 50 pA.